\documentclass[vecphys]{svmult}

\usepackage{makeidx}         % allows index generation
\usepackage{graphicx}        % standard LaTeX graphics tool
\usepackage{multicol}        % used for the two-column index
\usepackage[bottom]{footmisc}% places footnotes at page bottom
\makeindex             % used for the subject index
\begin{document}

\title*{The Stromlo Missing Satellites Survey}
% Use \titlerunning{Short Title} for an abbreviated version of
% your contribution title if the original one is too long
\author{Helmut Jerjen}
\authorrunning{Jerjen}
% Use \authorrunning{Short Title} for an abbreviated version of
% your contribution title if the original one is too long
\institute{Research School of Astronomy \& Astrophysics, Mt.~Stromlo Observatory,
Australian National University,
\texttt{jerjen@mso.anu.edu.au}}
%
%
% Use the package "url.sty" to avoid
% problems with special characters
% used in your e-mail or web address
%
\maketitle

\section*{Scientific Motivation}
\label{sec:1}
% Always give a unique label
% and use \ref{<label>} for cross-references
% and \cite{<label>} for bibliographic references
% use \sectionmark{}
% to alter or adjust the section heading in the running head
%
According to cosmological theory, density fluctuations of Cold Dark Matter (CDM)
form the first structures in the Universe. The gravitational potential wells of these 
dark matter halos suck in primordial gas and provide the seeds for the 
formation of stars via energy dissipation and cooling, a billion years after the Big Bang. 
The observational Universe today is filled with these galaxies, 
the prime repositories of shining baryonic matter. For obvious 
reasons, most of the detected and catalogued galaxies are intrinsically 
the largest and the brightest, those that can be seen from the greatest 
distance and are most easily studied against the night sky. Ironically, a 
major limitation on our ability to develop a consistent model 
that describes how galaxies emerged out of dark matter comes from the 
incompleteness of our picture of the nearby universe, in particular from 
the lack of a detailed understanding of the phenomenon {\it dwarf galaxies}. 

Dwarf galaxies are stellar systems composed almost entirely of dark matter
with a minimum mass of the order of $10^6$ solar masses. 
Examples have been discovered orbiting the Milky Way and Andromeda galaxies.
Extreme low star densities make them transparent and hard to find, 
but they seem to dominate by numbers any volume in space and were more 
numerous in the cosmological past. CDM theory tells us that the dark matter
mini halos and their optical manifestations, the dwarf galaxies, are the building blocks of larger galaxies like our Milky Way. 
Hence, if we want to shed light on the nature of dark matter and understand 
the driving mechanisms of galaxy formation/evolution we have to spend a disproportionate amount of effort on 
finding and physically characterising the faintest, most elusive galaxies 
that exist in the Universe.
%
%Thus if we want to make progress on existing 
%cosmological issues we have to spend a disproportionate amount of effort on finding 
%and physically characterising the faintest, most elusive stellar systems that exist in the Universe.
%
\vspace{-0.2cm}

\section*{Cold Dark Matter Theory on Galactic Scales}
A generic prediction of the standard CDM galaxy formation paradigm 
(e.g. Moore et al.~1999; Klypin et al.~1999; Governato et al.~2004) is that primary dark matter halos around massive galaxies contain hundreds of smaller clumps of dark matter. It is thought that the majority of these mini halos will gravitationally collect  sufficient primordial hydrogen gas and turn it into stars to form a dwarf satellite galaxy than can be observed today. However, in our best-studied case, the Milky Way, the predicted number of dark matter clumps exceeds that of the observed dwarf satellites by a factor of $\approx$20 (see Fig.~1). This current inconsistency between CDM cosmology and dwarf galaxy frequency is heavily debated in the literature and known as the {\it missing satellites} or {\it substructure problem} (Klypin et al. 1999; D'Onghia \& Lake 2004).

%
%
% For figures use
%
\begin{figure}[h]
\centering
% Use the relevant command for your figure-insertion program
% to insert the figure file.
% For example, with the option graphics use
\hspace*{-1cm}\includegraphics[height=9cm]{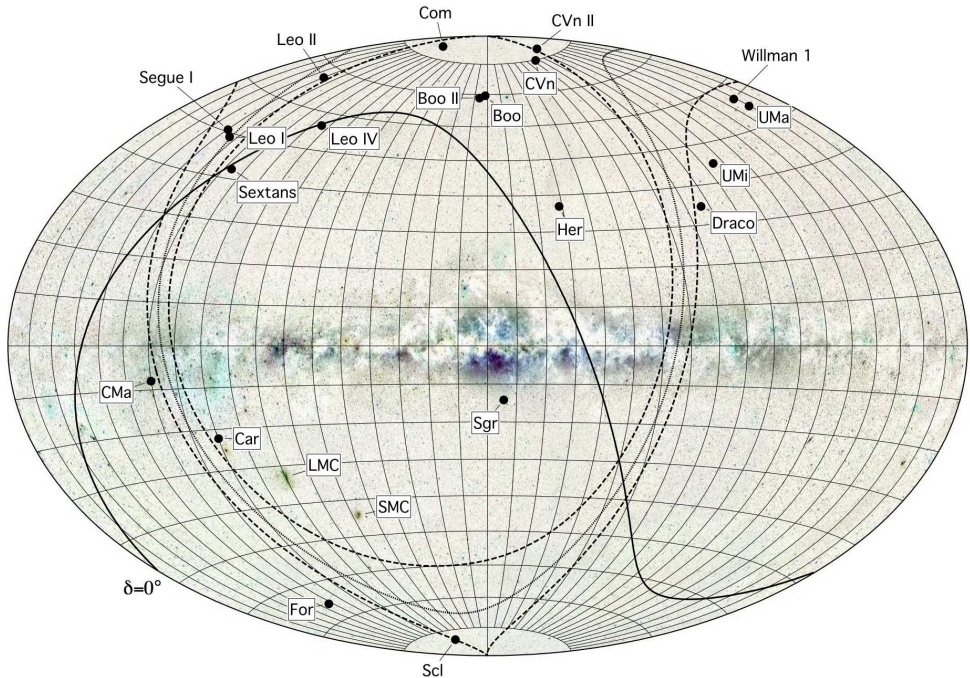}
\caption{Distribution of the known Milky Way satellites out to the Galactic virial radius 
$r_{vir}=250$\,kpc. Eleven dwarfs have been discovered in SDSS and 2MASS data in the last three years. The significance of the dwarfs being arranged in a plane (dotted line with the $\pm15^\circ$ edges
as dashed lines) has been discussed by Kroupa et al.~(2005) and Metz, Kroupa \& Jerjen (2007). Half of the entire sky (south of $\delta=0^\circ$) will be surveyed by the Stromlo Missing Satellites program.}
\label{fig:1}       % Give a unique label
\end{figure}

More recently the focus has shifted to whether the observed 3D-distribution of Milky 
Way satellites could actually be drawn from a population of dark matter subhaloes. 
The concern was raised when Kroupa et al.~(2005) and Metz, Kroupa \& Jerjen (2007) reported that the Milky Way satellites are statistically arranged in a disk, apparently
inconsistent with a cosmological substructure population at 
the 99.5\% confidence level. Although Kang et al.~(2005) argued that finding a 
dozen dwarf galaxies in a planar distribution is not improbable, debating this issue 
is futile as the incompleteness of the census of Milky Way satellites remains the 
ultimate uncertainty.   

Because incompleteness hinders any serious testing and possible refinement of the current cosmological model, progress can only be expected if observers 
can provide theoreticians with the full picture, including a robust dwarf satellite 
number and accurate estimates of their baryonic and dark matter contents, sizes, 
and galactocentric distances. The starting point of such a task is a deep and systematic 
photometric inventory of all the stars in the halo  of the Milky Way, a technical 
challenge that has become feasible just recently.

\section*{Previous Work}
{\bf Northern Hemisphere}: Willman et al.~(2005) conducted a first 
systematic blind search for Milky Way satellites using the Sloan Digital Sky Survey 
(SDSS; York et al.~2000) covering 25\% of the sky. Careful analyses of resolved 
stars in both the SDSS and the Two Micron All Sky Survey (2MASS)  revealed 
a first new Milky Way satellite, Ursa Major (UMa). Since then, 
nine more satellites have been reported (Zucker et al.~2006; Berlokurov et al.~2007;
Walsh, Jerjen \& Willman 2007).
%\medskip

\noindent {\bf Southern Hemisphere}: 
due to the lack of any digital imaging data until recently, the search for dwarf galaxies 
in the vicinity of the Milky Way generally had to rely on photographic plates
(e.g.~C\^ot\'e et al.~1998; Jerjen et al.~1998, 2000; Karachentsev et al.~2004). 
For example, Whiting et al.~(1999) detected the Cetus dwarf in the outskirts of the Local Group (at 780\,kpc) that was faintly visible on UK Schmidt plates.  Five years later, the great potential of finding new Milky Way satellites with modern technology was demonstrated when Martin et al.~(2004) discovered the 
Canis Major (CMa) dwarf in 2MASS.

\section*{The Stromlo Missing Satellites Survey}
The ANU\,1.35\,m SkyMapper telescope at Siding Spring Observatory
represents investment in Australian frontier technologies of A\$13 million. It is 
among the first of a new breed of specialised telescopes which are capable of 
scanning the sky more quickly and sensitively than ever before using a 
16k $\times$16k CCD mosaic camera with a 5.7 sq degree FOV (Keller et al.~2007; Tisserand et al., this volume). Over the next five years, that telescope is dedicated to carry out the multi-colour, multi-epoch Stromlo Southern Sky (S3) Survey generating 
150 Terabytes of CCD data. The final product will be a catalogue with positions and photometry for $\approx1$ billion objects in six bands: SDSS $u, g, r, i$, and $z$ plus an extra, Str\"omgren-like $v$ filter. The survey will cover all 20,000 square degrees south of the equator ($\delta<0^\circ$, see Fig.~1) and has photometric limits $0.5-1.0$\,mag fainter than SDSS. 

The various releases of the S3 catalogue will be systematically analysed by the Stromlo 
Missing Satellites (SMS) team employing sophisticated data mining algorithms that have
been developed and extensively tested with the publicly available Sloan DR4. 
Among others, we have announced the detection of  Bo\"otes\,II (Walsh, Jerjen \& Willman 2007; see also Walsh et al., this volume), a Milky Way satellite 
candidate with size-luminosity properties close to SEGUE\,1 (Belokurov et al.~2007), the second faintest of the currently known Milky Way companions.

Studies of such satellite candidates with a typical baryonic content of a few thousand stars or less require comprehensive imaging and spectroscopic follow-up programs to separate the wheat from the chaff. If the SMS project and the SDSS
survey, probing 75\% of the Milky Way's entire sphere of influence, will find 
significant numbers of true dwarf satellites that populate the same parameter space as 
model galaxies from high resolution simulations, these experiments would 
corroborate the standard model of cosmology in a remarkable way. Whatever 
the results, they will provide an unprecedented set of observational constraints 
that will show how tightly baryons and dark matter are bound on galactic scales
and will energise the debate about new physics.

%\printindex
\end{document}